\documentclass[doublecol]{epl2}

\title{Stabilization of silicon honeycomb chains\\ by trivalent adsorbates}
\shorttitle{Stabilization of HCC chains by trivalent adsorbates}
\author{C. Battaglia\inst{1} \and H. Cercellier\inst{1} \and C. Monney\inst{1} \and M.G. Garnier\inst{1}
\and P. Aebi\inst{1}} \shortauthor{C. Battaglia \etal}

\institute{
  \inst{1} Institut de Physique, Universit\'e de Neuch\^atel,
CH-2000 Neuch\^atel, Switzerland}


\abstract{The atomic structure of self-assembled
quasi-one-dimensional Gd chains on Si(111) has been investigated by
low-energy electron diffraction and scanning tunneling microscopy.
Based on comparison between Gd and Ca chains we show that this Gd
induced surface reconstruction belongs to the class of honeycomb
chain-channel structures. This clearly demonstrates that, besides
monovalent and divalent adsorbates, also trivalent adsorbates such
as Gd stabilize silicon honeycomb chains. Consequently silicon
honeycomb chains emerge as an universal building block in adsorbate
induced silicon surface reconstructions.}

\begin{document}

\maketitle

\section{Introduction}
\begin{figure}
\onefigure{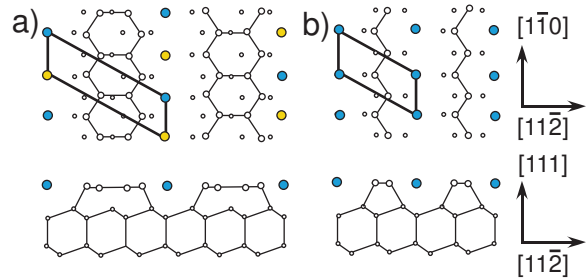}%
\caption{\label{fig:HCCSeiwatz}(Color online) a) Honeycomb chain and
b) Seiwatz chain model with adsorbates lying in the channels between
the chains. To stabilize the honeycomb chains, monovalent atoms are
required to occupy every site in the channels (blue and yellow
circles), whereas divalent adsorbates only occupy every second site
(blue circles only). Seiwatz chains are stabilized by divalent
adsorbates. Open circles are Si atoms. The (3$\times$1) and
(2$\times$1) unit cells are also shown.}
\end{figure}

Self-assembled atomic chains on silicon surfaces have been the
focus of intense research because of their quasi-one-dimensional
(1D) electronic properties and their interesting physics. Recently
the fluctuation and condensation phenomena at the metal insulator
phase transition of the In/Si(111) system could directly be
visualized via scanning tunneling microscopy (STM)
\cite{Lee05,Park05,Guo05,Ahn04}. Competing periodicities in
fractionally filled bands lead to the coexistence of different
Peierls distortions for the
gold induced reconstructions \cite{Crain03,Ahn05,Snijders06}.\\
Another important class of 1D systems are the alkali metal (AM=Li,
Na, K, Rb, Cs) and Ag induced, insulating (3$\times$1)
reconstructions formed by the deposition of 1/3 monolayer (ML)
onto the Si(111) surface. The AM/Si(111) systems adopt the
so-called honeycomb chain-channel (HCC) structure
\cite{Lottermoser98, Collazo-Davila98, Erwin98} shown in Fig.
\ref{fig:HCCSeiwatz}a) which is stabilized by the transfer of one
electron from the monovalent AM adsorbate into the Si surface
states.\\
A very similar reconstruction with (3$\times$"2") periodicity is
formed by adsorption of alkaline-earth metals (AEM=Mg, Ca, Sr,
Ba), where the $\times$"2" notation stands for a $\times$2
periodicity along the adsorbate chains but missing coherence
between adjacent chains \cite{Sakamoto02}. Due to the divalency of
the adsorbate only 1/6 ML, i.e. half the AM coverage, is required
to stabilize the HCC structure \cite{Lee01}. At 1/2 ML divalent
adsorbates induce a (2$\times$1) phase which was proposed to be
formed of $\pi$-bonded Seiwatz chains shown in Fig.
\ref{fig:HCCSeiwatz}b) \cite{Baski01,Sekiguchi01}. For
intermediate coverages, a series of 1D (n$\times$"2")
reconstructions, with n taking the values 5, 7 and even 9
depending on the adsorbate, is formed which are considered to be
composed of an appropriate combination of honeycomb chains and
Seiwatz chains (see Fig. \ref{fig:Model} for the 5$\times$"2"
case). Similar series of reconstructions were also observed for
the divalent rare earth metals (REM) Sm, Eu and Yb \cite{
Sakamoto05}. These REMs more commonly occur in the $3+$ valence
state, but depending on their chemical surrounding the $2+$
configuration is occasionally preferred as in this case. Thus up
to now, only monovalent and divalent adsorbates were found to
stabilize Si
reconstructions containing the honeycomb chain building block.\\
In this letter we focus on trivalent REMs, which exhibit chain
structures with (5$\times$"2") periodicity \textit{only}, but
whose detailed atomic structure has not been investigated.
Combining low-energy electron diffraction (LEED), STM and recent
angle-resolved photoemission spectroscopy (ARPES) results
\cite{Okuda04}, we show for the first time that the structure
induced by trivalent adsorbates contains the same honeycomb and
Seiwatz chains as in the chain reconstructions induced by divalent
adsorbates. The use of multiple complementary surface analysis
techniques is mandatory in the present case in order to derive a
reliable structural model. Based on electron counting we are also
able to explain, why only the (5$\times$"2") periodicity is
stabilized for trivalent adsorbates.
\section{Experiment}
We choose to investigate the Gd system, since it has recently been
demonstrated that predominantly single domain atomic Gd chains can
be grown on stepped Si(111) having a slight misscut of 1.1$^o$
towards the $[\bar{1}\bar{1}2]$ direction \cite{Kirakosian02}
allowing the use of macroscopic diffraction methods without domain
averaging. Qualitatively similar results are expected for the
observed (5$\times$"2") reconstruction induced by other trivalent
rare-earth metals Dy \cite{Engelhardt06},  Er \cite{Wetzel97} and Ho
\cite{Himpsel04}. Gd was evaporated from a water cooled e-beam
evaporator with of flux of $0.5\times10^{-4}$ ML/s at a pressure
below $5\times10^{-10}$ mbar onto the clean Si(111)-(7$\times7$)
substrate held at 680 $^o$C. The substrate was heated by passing a
direct current along the step direction $[1\bar{1}0]$. Growth and
experiments were carried out in an ultra high vacuum chamber with a
residual gas pressure of $3\times10^{-11}$ mbar equipped with an
Omicron LT-STM and Omicron Spectaleed LEED/Auger optics. For STM
measurements we used etched W tips.
\section{Results and Discussion}
\begin{figure}
\onefigure{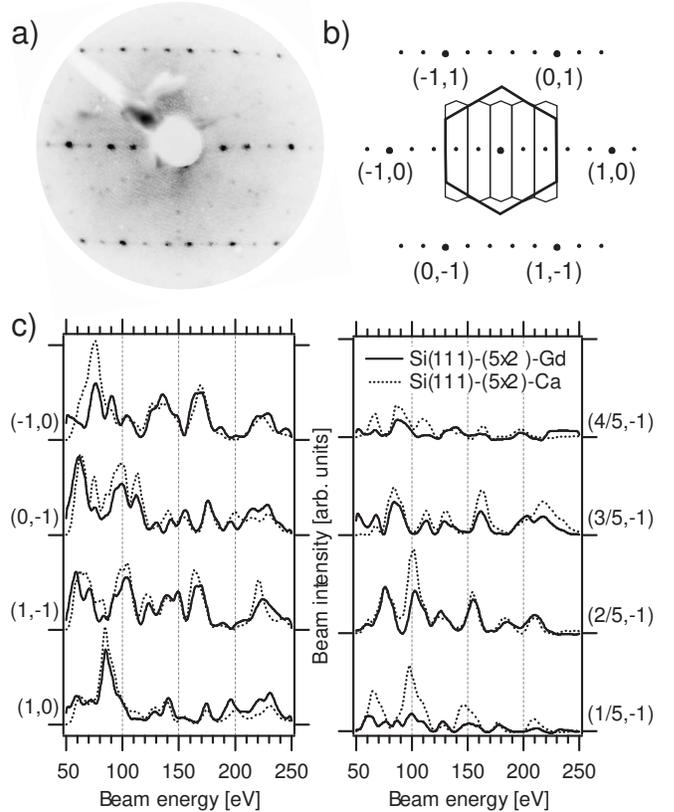}%
\caption{\label{fig:LEED2}a) LEED pattern of
Si(111)-(5$\times$"2")-Gd at 48 eV. b) Sketch of the 5$\times$1
LEED pattern in reciprocal space with beam indices. In real space
the chains are running along the vertical axis. c) Comparison
between experimental LEED IV curves of Si(111)-(5$\times$"2")-Gd
(full lines) and Si(111)-(5$\times$"2")-Ca (dotted lines).}
\end{figure}
Figure \ref{fig:LEED2}a) shows the LEED pattern of a typical
Si(111)-(5$\times$"2")-Gd surface with one dominant domain and
insignificant contributions from the two others and the
Si(111)-(7$\times$7) reconstruction. Only the (5$\times$1) spots
sketched in Fig. \ref{fig:LEED2}b) are clearly visible. The
$\times$2 periodicity along the chains manifests itself through
faint half-order streaks parallel to the $\times$5 spots (not
shown) observed only at certain energies. Similar streaks were
reported in studies of divalent adsorbate systems and explained in
terms of a stochastic distribution of adjacent chains with random
registry shifts leading to a (5$\times$"2") spot pattern with its
characteristic weak half-order streaks
\cite{Sakamoto02,Gallus02,Kuzmin05}.\\
Whereas the LEED spot positions only determine the type of Bravais
lattice of the  surface structure, i.e. its translational symmetry
properties, the point symmetries can be determined by a symmetry
analysis of the intensity vs voltage (IV) curves. The threefold
symmetry of the unreconstructed Si(111) surface termination is
broken by the growth of the chains. Whereas the (0,-1) and (-1,1)
beams are still equivalent as for the substrate, the (1,0) beam
exhibits a distinctive spectral signature as can be seen from Fig.
\ref{fig:LEED2}c). Thus only a mirror plane perpendicular to the chains is retained. \\
To obtain information about the atomic positions we compare LEED
IV curves from Si(111)-(5$\times$"2")-Gd to the curves from
Si(111)-(5$\times$"2")-Ca in Fig. \ref{fig:LEED2}c). IV-LEED
fingerprinting has played a crucial role in establishing the
equivalence between different AM induced (3$\times$1) HCC
reconstructions, since it was recognized that the
Si(111)-(3$\times$1)-AM reconstruction is predominantly a
substrate reconstruction with a common structure independent of
the adsorbate species \cite{Fan89}. Visual inspection of Fig.
\ref{fig:LEED2}c) already shows that the agreement between the Gd
induced and the Ca induced reconstruction containing one honeycomb
chain and one Seiwatz chain is surprisingly good. Most peak
positions of the Gd chains fall on the same energies as for the Ca
chains with comparable relative intensities. To obtain a
quantitative measure for the agreement between the two structures
we calculated Pendry's $R$ factor $R_p$ \cite{Pendry80}, which
takes into account the peak positions but also the relative
intensities between the peaks. For the integral order spots we
obtain $R_p=0.29$. For the fractional order beams we obtain
$R_p=0.35$. These values are similar to $R_p=0.36$ obtained by
Lottermoser \cite{Lottermoser98} comparing theoretical curves to
experimental data for the HCC model. The good agreement between
the two experiments suggests that both structures share the same
structural building blocks. Deviations may be due to the
difference in atomic number and the associated scattering
characteristics between Gd and Ca, differences in the precise adsorption geometry and coverage.\\
\begin{figure}
\onefigure{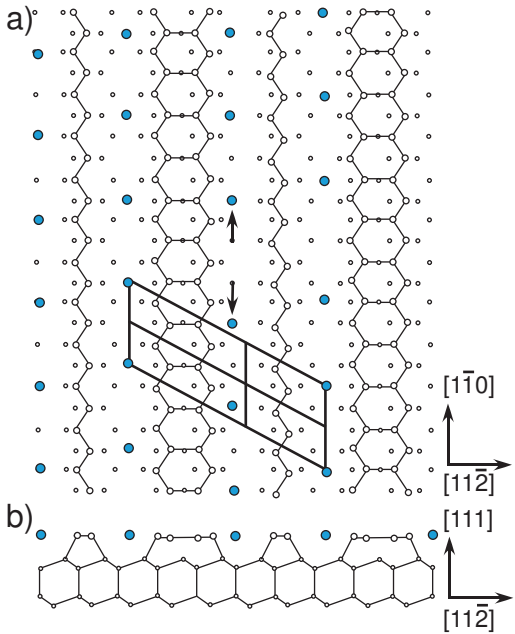}%
\caption{\label{fig:Model}(Color online) Structural model for the
Si(111)-(5$\times$"2")-Gd surface. Open circles are Si atoms, filled
blue circles are Gd atoms. The (5$\times$2) unit cell divided into
two (3$\times$1) and two (2$\times$1) unit cells is also shown.
Arrows indicate a registry shift of the adsorbates in the channel. }
\end{figure}
Adsorbate coverage is an important parameter for the determination
of any structural model. The exact amount of Gd at the surface is
difficult to determine accurately due to the fact that Gd diffuses
into the bulk above 600 $^o$C \cite{Kirakosian02}. The ideal
adsorbate coverage can however be determined when considering the
electron count required to stabilize the honeycomb and Seiwatz
chains. The HCC structure is known to be stabilized by the donation
of one electron per (3$\times$1) unit cell \cite{Erwin98,Lee01}.
Similarly the Seiwatz chain requires two electrons per (2$\times1$)
cell, since it may be stabilized by 1/2 ML of divalent adsorbates.
This is consistent with the number of surface states observed in
ARPES \cite{Sakamoto05b}. The (5$\times$2) unit cell can be thought
of as being build from two (3$\times$1) and two (2$\times$1) cells,
thus requires six electrons to be stabilized. Since Gd is trivalent,
the ideal coverage is two Gd atoms per (5$\times$2) cell or 1/5 ML.
This is in agreement with
the estimate of 0.2-0.4 ML given in Ref. \cite{Kirakosian02}.\\
The proposed  model for the Si(111)-(5$\times$"2")-Gd surface is
shown in Fig. \ref{fig:Model} consisting of alternating hexagonal
honeycomb chains and zig-zag Seiwatz chains made of Si. The
adsorbates are expected to form chains in the channels in between.
Due to the weak sensitivity of IV-LEED to the adsorbate itself, we
can not decide which absorption site is favored. Any structural
model must be consistent with results from other experimental
techniques. Fig. \ref{fig:STM} presents STM images of the Gd chains.
The overview a) shows long, parallel chains running along the
$[1\bar{1}0]$ direction. The separation between the rows is
consistent with the $\times$5 periodicity observed in LEED patterns.
High magnification empty and filled state images acquired in the
same scan to preserve their mutual registry are shown in Fig.
\ref{fig:STM}b) and c) respectively. The structural model is
superimposed. Based on simulated STM images derived from local
density approximation (LDA) calculations for the HCC structure
\cite{Erwin98,Lee01}, we identify the dark rows in the empty state
image with the location of the honeycomb chains. High intensity in
the empty state image is found along the adsorbate channels for both
the honeycomb and the Seiwatz chain structure \cite{Lee01,Jeong05}.
This is easily understood by noticing that the empty orbitals are
necessarily located on the adsorbate atom, since it donates its
electrons to the silicon chains. The filled state image c) appear as
triple rows of protrusions with $\times$2 periodicity along the
rows. The third row located along the Seiwatz chain (marked by S in
Fig. \ref{fig:STM}c) appears to lie slightly lower than the two main
rows (marked by H in Fig. \ref{fig:STM}c), which we identify with
the honeycomb chains.  In a previous STM study only the two main
rows H were resolved \cite{Kirakosian02}. The pairing of protrusions
causing the $\times$2 periodicity along the chains has been found to
be rather electronic in origin than geometric \cite{Lee03}. The
electrostatic attraction between a positive adsorbate ion and the
electrons in the neighboring saturated dangling bonds give rise to
such paired protrusions. We also remark that the registry of
neighboring chains is correct in our model. Careful inspection of
the filled state STM image shows that the honeycomb chain comes in
two configurations, either as two parallel rows of protrusions or in
a zig-zag configuration, indicated by empty circles in Fig.
\ref{fig:STM}c). Such a registry shift of only one period between
the two rows of the honeycomb chain is illustrated by the arrows in
Fig. \ref{fig:Model} and is simply due to a missing adsorbate and a
consecutive shift of all the following adsorbates by one period
along the chain direction. The local mixing of these two
arrangements with poor long range order is responsible for the
$\times$2 streaks seen in LEED patterns \cite{Gallus02}. Furthermore
this kind of defect leads to a local charge imbalance. It has been
suggested that additional Si adatoms are able to supply electrons
that dope the parent chain structure \cite{Erwin03} and may be able
to compensate for such missing charge. Additional Si atoms are
necessarily present since the formation of the HCC structure and
consequently also of the (5$\times$"2") structure is accompanied by
significant Si mass transport at the surface \cite{Saranin98} due to
the fact that the Si atom surface density of the (5$\times$"2")
structure is not equal to that of Si(111)-(7$\times$7). Although
steps may serve as a reservoir for reintegrating ejected Si atoms
into the surface, electromigration due to dc current heating
parallel to the steps does not favor the Si atoms to wander towards
the steps, resulting in a large number of randomly distributed
protrusions on
top of the chains. \\
\begin{figure}
\onefigure{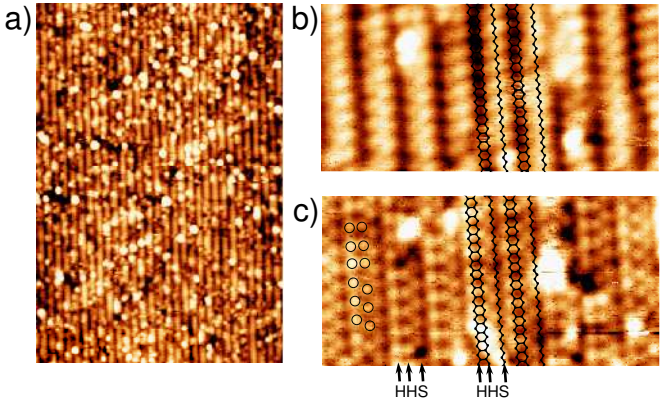}%
\caption{\label{fig:STM}(Color online) a) STM topography overview
(U= 1.9 V), 60 nm $\times$ 90 nm, b) and c) high resolution
topography of empty (U=1.9 V) and filled states (U=-1.9V), 18 nm
$\times$ 9 nm, I=0.18 nA. Arrows indicate the location of the
honeycomb chain (H) and Seiwatz chain (S). Empty circles mark the
two possible configurations of the honeycomb chain caused by a
registry shift of the Gd atoms in the adjacent channel as indicated
by arrows in Fig. \ref{fig:Model}.}
\end{figure}
We now turn to the discussion of recent ARPES results from
Si(111)-(5$\times$"2")-Gd \cite{Okuda04}, which provide additional
confirmation for our structural model. At least three semiconducting
surface states are observed at binding energies between 1 and 2 eV,
whose dispersions, band widths and symmetry properties are very
similar to those of the AM and AEM induced (3$\times$1) and
(3$\times$"2") reconstructions \cite{Okuda01} supporting a honeycomb
chain based structure. Furthermore ARPES data for the AEM induced
(5$\times$"2") structure resembles the one from the (3$\times$"2")
reconstruction. Very weak intensity is observed at the Fermi energy,
but has been interpreted as being due to defect states. A small
contribution to the spectral weight at the Fermi energy was also
observed in the semiconducting Si(111)-(3$\times$"2")-Ca system
\cite{Gallus02}, but was attributed to remaining (7$\times $7)
regions of pure silicon. Additionally, prolonged annealing of the Gd
induced reconstruction at 680 $^o$C leads to the nucleation of
metallic Gd silicide islands at the expense of the chain
reconstruction, which might possibly be responsible for the observed
photoelectron signal at the Fermi energy. However, from STM
measurements we do not find evidence for a metallic surface state
localized on the chains. We conclude that the Gd induced structure
is semiconducting and consequently requires an even number of
valence electrons per unit cell in agreement with the coverage of
two Gd atoms per (5$\times$2) unit cell. Therefore all ARPES results
fully support
our structural model.\\
A peculiar experimental finding to be explained is that Gd and
other trivalent REMs stabilize chain structures with the
(5$\times$"2") symmetry exclusively, whereas the monovalent
adsorbates stabilize only the genuine (3$\times$ 1) HCC structure
and the divalent adsorbates induce a series of (n$\times$"2")
reconstructions. Monovalent adsorbates must occupy every site
along the channel between the honeycomb chains to satisfy the
doping criterion. For lower coverages only parts of the
Si(111)-(7$\times$7) are transformed, whereas higher coverages
induce different surface structures. The stabilization of Seiwatz
chains requiring two electrons per unit cell is not possible.
Divalent adsorbates in turn must occupy every second site to
satisfy the doping balance. For higher coverages however,
additional adsorbates may be incorporated in the channels at the
expense of reducing every second honeycomb chain into a Seiwatz
chain. For trivalent adsorbates, charge balance requires that
every third site in the channel is occupied, if one wants to build
a structure exclusively formed by honeycomb chains. This is
apparently energetically unfavorable compared to an occupation of
every second site, which requires the combination of a honeycomb
chain with a Seiwatz chain resulting in the (5$\times$"2")
symmetry. The stabilization of a (5$\times$"2") period requires a
total of six electrons, a condition easily satisfied by taking two
trivalent adsorbates per unit cell. (7$\times$"2") and
(9$\times$"2") reconstructions are not observed for the trivalent
adsorbates. Consisting of one honeycomb chain and two respectively
three Seiwatz chains, they require 10 respectively 14 electrons
per unit to be stabilized, a condition which can not be satisfied
by trivalent donors. Electron counting thus provides a simple
intuitive picture for the occurrence of the various phases.
\section{Conclusion}
Driven by the elimination of dangling bonds and relief of surface
stress, silicon surfaces reconstruct in strikingly diverse ways.
Among the large variety of adsorbate induced reconstructions, the
honeycomb chain emerges as a most stable building block allowing
maximum reduction of the surface energy. The fact that only
silicon atoms participate in the formation of the honeycomb chains
allows a variety of adsorbates to adopt the HCC structure by
donating the correct number of electrons to the substrate.
Combining the complementary strength of IV-LEED fingerprinting,
STM and ARPES, we demonstrated for the first time that next to
monovalent and divalent adsorbates, \emph{trivalent adsorbates are
also able to stabilize the honeycomb chains}. Based on a intuitive
electron counting model, we are further able to explain, why only
the (5$\times$"2") symmetry is stabilized by trivalent adsorbates.
Our conclusions allow to enlarge the range of honeycomb chain
stabilizing adsorbates to the trivalent elements.

\bibliography{GdSi111}

\acknowledgments Helpful conversations with Celia Rogero, Laurent
Despont, Christian Koitzsch and Jos\'{e} A. Martin-Gago are
gratefully acknowledged. Skillfull technical assistance was provided
by our workshop and electric engineering team. This work was
supported by the Fonds National Suisse pour la Recherche
Scientifique through Div. II and MaNEP.

\end{document}